\documentclass[conference]{IEEEtran}
\IEEEoverridecommandlockouts
\let\oldvec\vec
\usepackage{xcolor}
\usepackage{breqn}
\usepackage{multirow}
\usepackage{float}
\usepackage[compact]{titlesec}
\titlespacing{\section}{10pt}{3pt}{3pt}
\usepackage{relsize}
\usepackage{interval}
\usepackage[linesnumbered,ruled,vlined]{algorithm2e}
\usepackage{tabularx}
\usepackage{multirow}
\usepackage{amsmath,etoolbox}
\usepackage[ruled,vlined]{algorithm2e}
\usepackage{cite}
\usepackage{amsmath,amssymb,amsfonts}
\usepackage{algorithmic}
\usepackage{graphicx}
\usepackage{textcomp}
\usepackage{xcolor}
\usepackage{enumitem}
\setlist{nolistsep,leftmargin=.6cm}
\def\BibTeX{{\rm B\kern-.05em{\sc i\kern-.025em b}\kern-.08em
    T\kern-.1667em\lower.7ex\hbox{E}\kern-.125emX}}
\usepackage{amsmath, amssymb, bm, cite, epsfig, psfrag}
\usepackage{graphicx}
\usepackage{cuted}
\usepackage{soul} 
\usepackage[left=0.65in, right=0.65in, bottom=0.99in, top=0.66in]{geometry}
\usepackage{dblfloatfix}
\usepackage{array}
\usepackage{lipsum}
\newcolumntype{P}[1]{>{\centering\hspace{0pt}}p{#1}}
\newcolumntype{M}[1]{>{\centering\hspace{0pt}}m{#1}}
\newcolumntype{L}{>{\centering\arraybackslash}m{3cm}}
\usepackage[font=small]{caption}
\usepackage{epstopdf}	
\usepackage{longtable}
\usepackage{supertabular,booktabs}
\usepackage{bbm}
\usepackage{multirow}
\renewcommand{\arraystretch}{1.5}
\usepackage{subfigure}
\usepackage{etoolbox}
\usepackage{pbox}
\usepackage{fixltx2e}
\usepackage{tabu}	
\usepackage{filecontents}
\usepackage{enumerate}
\usepackage{textcomp}

\usepackage{colortbl}
\usepackage{fancyhdr}

\usepackage{bm}
\newtoggle{conference}
\togglefalse{conference} 
\def\argmin{\mathop{\mathrm{arg\,min}}}

\makeatletter
\newsavebox\myboxA
\newsavebox\myboxB
\newlength\mylenA
\newcommand*\xoverline[2][0.75]{%
    \sbox{\myboxA}{$\m@th#2$}%
    \setbox\myboxB\null
    \ht\myboxB=\ht\myboxA%
    \dp\myboxB=\dp\myboxA%
    \wd\myboxB=#1\wd\myboxA
    \sbox\myboxB{$\m@th\overline{\copy\myboxB}$}
    \setlength\mylenA{\the\wd\myboxA}
    \addtolength\mylenA{-\the\wd\myboxB}%
    \ifdim\wd\myboxB<\wd\myboxA%
       \rlap{\hskip 0.5\mylenA\usebox\myboxB}{\usebox\myboxA}%
    \else
        \hskip -0.5\mylenA\rlap{\usebox\myboxA}{\hskip 0.5\mylenA\usebox\myboxB}%
    \fi}

\newtheorem{theorem}{Theorem}

\begin{document}

\title{Access Delay Constrained  Activity Detection in Massive Random Access}

\author{\IEEEauthorblockN{Jyotish Robin, Elza Erkip}
\IEEEauthorblockA{\textit{Dept. of Electrical and Computer Engineering,}\\
\textit{Tandon School of Engineering, New York University, Brooklyn, NY, USA}}
}

\maketitle
\begin{abstract}
In 5G and future generation wireless systems, massive IoT networks with bursty traffic are expected to co-exist with cellular systems to serve  several latency-critical applications. Thus, it is important for the access points to identify the active devices promptly with minimal resource consumption to enable massive machine-type communication without disrupting the conventional traffic. In this paper,  a frequency-multiplexed strategy based on group testing is proposed for activity detection which can take into account the constraints on network latency while minimizing the overall resource utilization. The core idea is that during each time-slot of active device discovery, multiple subcarriers in frequency domain can be used to launch group tests in parallel to reduce delay. Our proposed scheme is functional in the asymptotic and non-asymptotic regime of the total number of devices $(n)$ and the number of concurrently active devices $(k)$.  We prove that, asymptotically, when the number of available time-slots scale as $\Omega\big(\log (\frac{n}{k})\big)$, the frequency-multiplexed group testing strategy requires $O\big(k\log (\frac{n}{k})\big)$ time-frequency resources which is order-optimal and results in an $O(k)$ reduction in the number of time-slots with respect to the optimal strategy of fully-adaptive  generalized binary splitting. Furthermore, we establish that the frequency-multiplexed GT strategy shows significant tolerance to estimation errors in  $k$. Comparison with 3GPP standardized random access protocol for NB-IoT  indicates the superiority of our proposed strategy in terms of access delay and overall resource utilization.  
\end{abstract}

\begin{IEEEkeywords}
  IoT, low latency, active device discovery, group testing, massive random access.
\end{IEEEkeywords}

\section{Introduction}~\label{sec:intro}
With the inception of massive IoT networks comprising of millions of smart devices, cellular systems are expected to support Machine-To-Machine traffic in addition to the conventional Human-to-Human traffic to take advantage of the widespread coverage\cite{7823342}. However, Machine-to-Machine traffic is typically characterized by bursty activity pattern, access delay constraints, limited power budget etc. which are in significant contrast to the traditional cellular systems. Therefore, an effective adaptation of cellular networks to handle Machine Type Communication (MTC) calls for revision of various protocols being used in the Radio Access Network.

Narrowband-Internet of Things (NB-IoT) and LTE-MTC are two popular 3GPP proposed standards to provide connectivity for MTC \cite{8422799}. For in-band operation within the bandwidth of LTE carrier,
NB-IoT uses a minimum system bandwidth of 180 kHz. In the standardized design of NB-IoT physical random access channel (NPRACH), within the OFDM resource  grid of 180 kHz bandwidth, an NPRACH band consisting of 12, 24, 36, or 48 subcarriers can be configured. NB- IoT is  expected to serve applications requiring low data rates with a latency of 10s or less. On the other hand, LTE-MTC is optimized for applications with higher data rates and lower latency.

In Random Access (RA) procedure, the first phase is typically detection of the active devices so that resources can be allocated for upcoming data transmissions. RA in an NB-IoT system is a 4-stage handshake protocol in which the device that needs to establish a connection with the Access Point (AP) randomly chooses and transmits a preamble \cite{8288195}. The preamble is
distinctively identified by its frequency hopping pattern. In each RA attempt, based on the Coverage Extension (CE) level of the device, the preamble is repeated a certain number of times. If multiple devices from a  CE level choose an identical preamble during the same PRACH instance, it leads to a collision.
The frequency of preamble collisions sharply rise when there is a massive number of devices trying to access the network.

In \cite{7823342}, the authors reviewed several  proposals to tackle the RA congestion problem in modern cellular systems. Compressive sensing based techniques for activity discovery exploiting the sporadic and sparse nature of sensor activities have been investigated in \cite{Ke_2020,8323218,8227652}. Specifically, \cite{8227652} compares compressed-sensing  and coded
slotted ALOHA in terms
of user activity detection performance vs resource utilization.

Recently, there has been an upswing of research interest in Group Testing (GT) based designs for active device identification in massive random access \cite{8262800, robin2021sparse,8926588}. The basic principle is as  follows: Each device is assigned a signature sequence of 0's and 1's. During active device discovery, each sensor uses  on-off signaling  to transmit their signature and at the AP, an energy detector is used. This leads to a Boolean-OR channel model and the AP identifies the active devices based on the channel outputs. In \cite{8262800}, Inan et al. proposed random access protocols for massive access based on non-adaptive GT. In our recent work\cite{robin2021sparse}, we considered energy constrained active device discovery using GT when there are multiple clusters of devices with different activity patterns. None of this prior literature explicitly considers the availability of frequency resources in designing GT schemes.

In this paper, we use GT techniques to address the following critical aspects of random access in massive-cellular IoT by utilizing both time and frequency resources.

\noindent \textbf{i) Latency}: Many MTC applications are delay sensitive as the devices typically carry information requiring timely delivery. 
\\ \noindent \textbf{ii) Resource utilization}: The MTC-specific RA schemes should ensure that  the massive density of devices does not deplete the resources available for Human-to-Human traffic. Thus, minimizing resource utilizition during active device discovery is paramount for efficient accomodation of MTC. 

If there are no access delay constraints, the optimal strategy to minimize the total resource utilization is based on Hwang's generalized binary splitting (GBS) \cite{doi:10.1142/1936}. On a high level, this involves applying the binary splitting algorithm recursively to the $k$ groups of size approximately $n/k$ where $n$ is the total number of devices out of which $k$ are active. 
For sufficiently large $n/k$, the number of time-slots required by GBS is approximately  $ k\log (\frac{n}{k})$. Though optimal in terms of overall resource utilization, the adaptive nature of GBS leads to potential violation of the access delay constraints since there is no room for parallelization in frequency (except when $n <2k-1$) as each step of the algorithm is adaptively designed based on the previous steps.

In our work, we propose a strategy based on Li's method for screening experimental variables \cite{10.2307/2281652} which can take into account the access delay constraints. The core idea is that in a given time-slot, multiple subcarriers are available as stated in \cite{3gpp.36.331} to launch group tests in parallel thereby potentially reducing the overall access delay. Our strategy is applicable in the asymptotic as well as non-asymptotic regime of $n$ and $k$. In the asymptotic regime, the results indicate that the worst-case resource utilization for our proposed frequency-multiplexed GT strategy can be $O\big(k\log (\frac{n}{k})\big)$ which is order-optimal while reducing the access delay by a factor of $O(k)$.

The remainder of this paper is organized as follows. In
Section II we formulate the problem of active device discovery using group testing. In Section III, we propose the frequency-multiplexed GT strategy for activity detection and discuss how it operates. The impact of estimation errors in $k$ on the performance is also characterized. Section IV analyses the feedback requirements inherent to the various active device discovery  strategies we discussed. Section V presents numerical simulations. Finally, Section VI concludes the paper.
\section{Problem Formulation}~\label{Sysmodel}
Let $\mathcal{U}=\{s_{1},s_{2}, ...,s_{n}\}$ represent the set of devices with cardinality $|\mathcal{U}|=n$. Each device $s_i \in \mathcal{U}$ can be in one of the two possible states, viz, Active State $(\mathcal{AS})$ and Inactive State $(\mathcal{IS})$ independent of the state of other devices. Among the $n$ devices,  $k$  are active and we assume that the value of $k$ is known beforehand. We will discuss the impact of not knowing $k$ in Section III.

We use a GT based approach for detecting the set of active devices. A group testing matrix W is defined as a binary matrix formed by a  set of  $n$-coordinate column vectors, $\mathbf{w}_{\ell} \in\{0,1\}^{n}$ where, $\ell \in\{1,2, \ldots, L\}$. i.e.,
\begin{equation}
    \mathbf{W}=\left[\mathbf{w}_{1}, \ldots, \mathbf{w}_{L}\right]=\left[\mathbf{x}_{1}, \ldots, \mathbf{x}_{n}\right]^{\intercal} \in \mathbb\{0,1\}^{n \times L}
    \label{GTmatrix}
\end{equation} where the $i^{th}$ row is a binary signature of length $L$ designed for the $i^{th}$ device. In the active device discovery phase, each active device transmits its  binary signature (On-Off keying) in a time-synchronized manner over the $L$ probes. Each probe potentially involves  a group of devices transmitting at the same time over some frequency band if there are multiple active devices with a $\textbf{1}$ at identical indices in their signatures. Also, using different subcarriers in frequency domain, multiple probes can be initiated during the same time-slot.  In massive access scenarios, acquiring each device's channel state information is infeasible as it typically needs an overwhelming amount of pilot resources. Thus, we assume that, at the AP,  a non-coherent energy detector is used to  make a binary decision indicating the presence of energy in the received signal. This operation does not require any CSI.  Let ${\mathbf{y}}=\left(y_{\ell}\right) \in\{0,1\}^{L}$  indicate the results  vector. 
\begin{equation}
y_{\ell}=\left\{\begin{array}{ll}
1 & \text { if energy detected }\left(\exists i \in \mathcal{A S} \text { with } \mathbf{w}_{\ell}(i)=1\right) \\
0 & \text { if no energy detected }\left(\forall i \in \mathcal{A S}, \mathbf{w}_{\ell}(i)=0\right)
\end{array}\right.
\end{equation}
 The $\ell^{th}$ probe is a positive probe if $y_\ell=1$ and a negative probe if $y_\ell=0$. We assume that there are no errors in detecting the energy. Given the matrix $\textbf{W}$ composed of binary signatures and results vector $({\textbf{y}}$), we need to identify the active devices efficiently. This  model is equivalent to the GT problem considered  in \cite{8262800}.
 
 During active device identification phase, we are bound to operate within a time-frequency resource grid  $ \mathcal{G}:[N_{f}^{UB} \times N_t^{UB}]$ determined by bandwidth and access delay constraints. In effect, the $L$ probes need to be efficiently distributed across multiple time-slots and multiple subcarriers. Here, $N_f^{UB}$ is  the upper bound on the number of subcarriers available for active device discovery and $N_t^{UB}$ is the upper bound on the number of available time-slots determined based on the access delay requirements of the devices. 

As per 3GPP proposals to alleviate the RA congestion problem in massive access, the frequency domain RACH resources can be dynamically varied\cite{7823342,3gpp.36.331}. Hence, it is reasonable to assume that $N_f^{UB}$ is large enough and we typically operate well within as we show in Table I. 

We use $N_f(i)$ to denote the number of parallel subcarriers needed in the $i^{th}$ time slot. Thus, the total resource utilization is $N_R=\sum _{i=1}^{N_t}N_f(i).$ Our aim is to minimize the worst-case total resource utilization required for identifying the active devices using GT while operating within the access delay constraint  $N_t \leq N_t^{UB}$ where  $N_t$ denotes the number of time-slots we use. We will assume availability of feedback after every time-slot. In Section IV, we comment on the feedback requirements for various schemes discussed in this paper. 

Our modeling doesn't have any sparse activity assumptions. However, the bounds we establish gets tighter if $k<<n.$

 \section{Frequency-multiplexed GT for Access delay constrained activity detection}

\begin{figure}   
	\centering
	\includegraphics[width=0.75\linewidth,height=50mm]{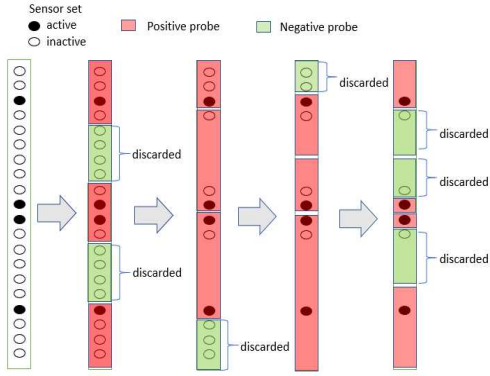}
	\setlength{\belowcaptionskip}{-10pt}
	\caption{Frequency-multiplexed GT strategy for activity detection.}
	\label{fig:Parallelization}
	\end{figure}

 Motivated by Li's work in \cite{10.2307/2281652}, we consider a frequency-multiplexed GT strategy in which multiple disjoint groups of devices are probed using different subcarriers during each time-slot. Thus, at each time-slot, no device is subjected to probing more than once. Moreover, we assume that the group sizes remain fixed during each time-slot. After each time-slot, based on the probing results, we eliminate the devices which are certainly inactive and disjointly partition the reduced set of devices in the further iterations. A sample illustration is shown in Fig. \ref{fig:Parallelization} where the parameters are $n=20$ and $k=4$.

Assume that the devices are partitioned into $G_i$  groups of size $S_i$ during the $i^{th}$
time slot. Note that $G_i$ also corresponds to the number of frequency resources used in time-slot $i$. After the $i^{th}$ time-slot, at most $k$ groups can result in positive probes and therefore at most $kS_i$ devices need further evaluation.   Thus, the worst case resource utilization ($N_R$) can be written as:\vspace{-0.1cm}
\begin{equation} \vspace{-0.1cm}
    N_R = \frac{n}{S_1}+\frac{kS_1}{S_2}+\ldots +kS_{N_t^{UB}-1}.
    \label{mneq}
\end{equation}
Though $S_i$'s in (\ref{mneq}) are discrete variables, we consider its differentiable interpolation for the purpose of analysis.
Taking derivatives w.r.t $S_i$ $\forall i \in \{1,2,\ldots,N_t^{UB}\}$, the RHS in (\ref{mneq}) can be minimized to obtain the following optimal parameters: 
\begin{equation}
    S_i^*= \bigg(\frac{n}{k}\bigg)^{\frac{N_t^{UB}-i}{N_t^{UB}}}
    \label{subsz}
\end{equation}
\vspace{-0.3cm}
\begin{equation}
    G_i^*= k \times \bigg(\frac{n}{k}\bigg)^{\frac{1}{N_t^{UB}}}.
    \label{gsiz}\vspace{-0.2cm}
\end{equation}\vspace{-0.1cm}Using (\ref{subsz}) and (\ref{gsiz}) in (\ref{mneq}), we get \vspace{-0.1cm}  \begin{equation}
    N_R^*=k \times N_t^{UB} \times \bigg(\frac{n}{k}\bigg)^{\frac{1}{N_t^{UB}}}
    \label{nr*}\vspace{-0.1cm}
\end{equation}

 The values $S_i^*, G_i^*$ \& $N_R^*$ need to be appropriately rounded off to an integer. Note that (\ref{gsiz}) implies that the optimal frequency-multiplexed GT strategy involves keeping approximately the same number of groups (frequency resources) during each time slot $i \in \{1,2,\ldots,N_t^{UB}\}$. 
    \vspace{-0.1cm}
\subsection{Optimal number of time-slots}

Eqn. (\ref{nr*}) indicates that $ N_R^*$ is a convex function of a continuous interpolation of $N_t^{UB}$. Thus, though we are allowed to use upto $N_t^{UB}$ time-slots, the optimal choice for $N_t$  can be less than $N_t^{UB}$ from a total resource utilization viewpoint. Taking derivative of $N_R=k \times N_t \times \Big(\frac{n}{k}\Big)^{\frac{1}{N_t}}$  w.r.t $N_t$ gives \vspace{-0.2cm}
\begin{equation} \vspace{-0.2cm}
    N_t^{opt}\approx \ln \left(\frac{n}{k}\right).
\end{equation}
Now, using  $N_t^{opt}$ in (\ref{gsiz}) and (\ref{nr*}), we have the following results.

\vspace{0.15cm}
\noindent \textit{Theorem 1:}
The minimum resource utilization $(N_R^{min})$ for discovering $k$ active devices from a population of size $n$  using  frequency-multiplexed GT is upper-bounded as follows:
\small
\begin{equation}
    \text{1) If } N_t^{UB} \geq \left \lceil \ln \left(\frac{n}{k}\right)\right \rceil :N_R^{min}\leq e k\times \left \lceil \ln \left(\frac{n}{k}\right)\right \rceil. \hspace{1cm}
    \label{nrmin}
\end{equation}
\begin{equation}
    \text{2) If } N_t^{UB} = \alpha  \left \lceil \ln \left(\frac{n}{k}\right)\right \rceil  :N_R^{min}\leq \alpha k  e^{1/\alpha} \times \left \lceil \ln \left(\frac{n}{k}\right)\right \rceil \hspace{0.2cm}
    \label{nrmin2}
\end{equation}
\normalsize where $\alpha \leq 1$.
\vspace{0.2cm}

\noindent \textbf{Remark 1.} Note that the upper bound in (\ref{nrmin}) for minimum resource utilization using the frequency-multiplexed GT strategy exceeds the resource utilization of Hwang's GBS strategy \cite{doi:10.1142/1936} approximately by a factor of $\frac{e}{\log e}$. \vspace{0.1cm} \\
\textbf{Remark 2.} Asymptotically, if $N_t^{UB}$ scales as $\Omega \big(\log \left(\frac{n}{k}\right)\big)$, the frequency-multiplexed GT strategy is order-optimal as its resource utilization is $O(k\log \left(\frac{n}{k}\right)).$ Thus, there is an $O(k)$ saving in time-slots when compared to the Hwang's GBS.
\vspace{0.15cm}

\noindent \textit{Corollary 1:}
The optimal number of  subcarriers $(G_i^{opt})$ required for active device identification using the frequency-multiplexed GT strategy is as follows:
\small \begin{equation}
    \text{1) If } N_t^{UB} \geq \left \lceil \ln \left(\frac{n}{k}\right)\right \rceil :G_i^{opt}\approx e \times k \hspace{1.6cm}
    \label{corol}
\end{equation}
\begin{equation}
    \text{2) If } N_t^{UB} = \alpha  \left \lceil \ln \left(\frac{n}{k}\right)\right \rceil  :G_i^{opt}\approx e^{1/\alpha} \times k \hspace{0.9cm}
    \label{corol1}
\end{equation} 
\normalsize

In Table \ref{tab:tab1}, the number of time-slots and subcarriers per time-slot used by our proposed strategy is shown for several values of $n,k$ and $N_t^{UB}.$ We now compare these values with the 3GPP standardized preamble based random access for NB-IoT. In NB-IOT, one NPRACH preamble consists of four symbol groups, with each symbol group comprising of one CP and five OFDM symbols \cite{7876968}. The duration of NPRACH preamble is 5.6 ms or 6 ms. The probability that, in a given RA opportunity, a device completes preamble transmission without collision when $s$ NPRACH subcarriers are available  as given in \cite{s19143237} is\vspace{-0.1cm}
\begin{equation} \vspace{-0.1cm}
    \mathcal{P}_{k, s}=1-\left(1-1/s\right)^{k-1}, \quad k>0 \label{psucpr}
\end{equation} where $s \in \{12,24,36,48\}.$ Note that in (\ref{psucpr}) only collisions, and no noise is considered. Since, a symbol group is the atomic unit of NPRACH preamble, we consider the number of time-slots required for an NPRACH preamble  as $N_t^{IOT}=4.$ Assuming $k=5$ active devices  and $s=48$ subcarriers in (\ref{psucpr}) leads to a success probability of 92\%. Furthermore, to achieve a success probability greater than 99\%, two repetitions of NPRACH preamble is required leading to 
$N_t^{IOT}=8$. In contrast, as shown in Table \ref{tab:tab1}, with $N_t^{UB}=8$ time-slots, only $G_i^{opt}=14$ subcarriers are required by our proposed strategy which corresponds to around 70\%  bandwidth savings.
\begin{table}
\begin{tabular}{|c|l|c|c|c|c|c|c|c|}
\hline {  } & \multicolumn{2}{|c|}{$n=10^{3}$} & \multicolumn{2}{|c|}{$n=10^{3}$} & \multicolumn{2}{|c|}{$n=10^{4}$} & \multicolumn{2}{|c|}{$n=10^{4}$} \\$ \text{}N_t^{UB}$ & \multicolumn{2}{|c|}{$k=5$} & \multicolumn{2}{|c|}{$k=10$}& \multicolumn{2}{|c|}{$k=5$} & \multicolumn{2}{|c|}{$k=10$} \\
\cline { 2 - 9 } & $G_i^{opt}$&$N_t$ & $G_i^{opt}$&$N_t$& $G_i^{opt}$&$N_t$ & $G_i^{opt}$&$N_t$ \\
\hline 2 & 71 & 2 & 101& 2 & 224& 2 & 317 & 2 \\
\hline 4 & 19 & 4 & 32 & 4 & 34 & 4 & 57 & 4 \\
\hline 6 & 14 & 6 & 28 & 5 & 18 & 6 &  32 & 6 \\
\hline 8 & 14 & 6 & 28 & 5 & 14 & 8 &  28 & 7 \\
\hline
\end{tabular}
\caption{\label{tab:tab1} Number of subcarriers ($G_i^{opt}$) and time-slots $(N_t)$   used by frequency-multiplexed GT for a given $n,k$ and $N_t^{UB}$.} \vspace{-0.3cm}
\end{table} 
\subsection{What happens if $k$ is unknown?}	

In a practical scenario, it may not be possible to precisely know the number of active devices,  $k$. Thus, we have to rely on estimates of $k$ and hence it is important to understand how sensitive is the resource utilization $(N_R)$ to an estimation error in $k$. Assume $\hat{k}$ is an estimate of $k$. In this case, the scheme works as follows: Firstly, the subgroup sizes are computed using $\hat{k}$ in (\ref{subsz}) as\vspace{-0.2cm} \begin{equation}
    S_i^*(\hat{k})= \bigg(\frac{n}{\hat{k}}\bigg)^{\frac{N_t-i}{N_t}}, \text{ where } N_t=min(N_t^{UB},N_t^{opt}).
    \label{sopt}\vspace{-0.2cm}
\end{equation}  As indicated in  (\ref{mneq}), during the first time-slot  $\frac{n}{S_1}=\hat{k}  \Big(\frac{n}{\hat{k}}\Big)^{\frac{1}{N_t}}$ subcarriers are needed. After any time-slot, at most $k$ probes can be positive probes. Thus, at most $k \times S_i$ devices need further probing after the $i^{th}$ time-slot, for which the devices are divided disjointly into groups of size $S_{i+1}$. Hence  $\frac{kS_i}{S_{i+1}}={k}  \Big(\frac{n}{\hat{k}}\Big)^{\frac{1}{N_t}}$ subcarriers are required during any time-slot other than the first time-slot. 
Thus, the worst case overall resource utilization in this case is given by \vspace{-0.25cm} \begin{equation}
     N_R^*(\beta)=k(N_t+\beta-1) \Bigg(\frac{n}{\beta k}\Bigg)^{\frac{1}{N_t}}.
     \label{betares}\vspace{-0.2cm}
\end{equation} Note that $\beta=1$ corresponds to zero estimation error. Therefore, the percentage increase in the resource utilization $(J)$ when we use $\hat{k}$ instead of $k$ is given by,
\begin{equation} \vspace{-0.1cm}
    J(\beta)=\frac{N_R^*(\beta)-N_R^*(1)}{N_R^*(1)}=\frac{N_t+\beta-1}{N_t} \times \Big(\frac{1}{\beta}\Big)^{\frac{1}{N_t}}-1.
\end{equation}

\begin{figure}   
	\centering
	\includegraphics[width=0.6\linewidth,height=45mm]{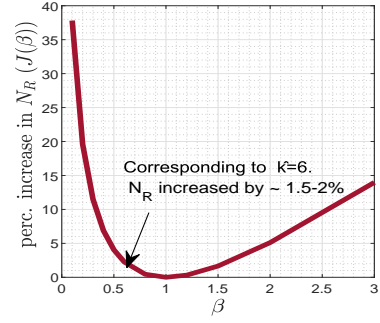}
	\setlength{\belowcaptionskip}{-10pt}
	\caption{Percentage increase in resource utilization, $J(\beta)$ due to estimation errors in $k$. $\beta=\hat{k}/k$. Assume $n=1000, k=10, N_t^{UB}=4.$ }
	\label{fig:esterror}
	\end{figure}

As an example, consider the case of $n=10^3$, $k=10$  and $N_t^{UB}=4$  as shown in Table \ref{tab:tab1}. Since $\ln(100)>4$, we have $N_t=4$. If $k$ is known a priori we require 32 subcarriers per time-slots leading to a worst-case resource utilization of $N_R=128$.. Now, assume that $k$ is unknown and we only have an estimate $\hat{k}=6.$ From (\ref{sopt}),  in the first time-slot, we use $6 \times (1000/6)^{1/4}\approx 22$ subcarriers. The number of subcarriers needed in the other time-slots equals $ 10 \times   (1000/6)^{1/4} \approx 36$ . This corresponds to a worst-case overall resource utilization of 130. Notably, the excess resources required is around 1.5\% as shown in Figure  \ref{fig:esterror}. From Figure \ref{fig:esterror}, one can note that,  as the estimation error increases in either direction, the resource utilization also increases. However, it does not change significantly even with reasonable estimation errors. 
\section{Feedback considerations}
We assume that the AP coordinates different devices after each time-slot so that the devices will know which probing group they belong to. Thus, after each time-slot, there is a need for feedback to assist the design of further stages. In contrast, with non-adaptive GT-based algorithms, the probing schedule is predefined and all the devices are aware of this schedule beforehand. However, the non-adaptive strategies require $O(k^{2} \log n )$ time-frequency resources for successful detection of active devices which is significantly larger than the resource utilization of adaptive algorithms we considered.  In this section, we evaluate the feedback overheads pertaining to the different adaptive strategies.

Owing to the fully adaptive nature of Hwang's GBS, 1-bit of feedback needs to be transmitted by AP after each of the $k \log \big(\frac{n}{k}\big)$ (approximately) time-slots to convey the  probing result. The total feedback required in this case is approximately $k \log \big(\frac{n}{k}\big)$ bits.

Next, we will derive $N_{FB}$, the amount of feedback required (in bits) for our proposed frequency-multiplexed GT.

\hspace{-0.2cm}\textbf{Case 1: }$N_t^{UB} \geq N_t^{opt}$  

As established in Corollary 1, the optimal number of groups at each time-slot is given by $G_i^{opt} \approx e \times k.$ At each of the $N_t^{opt}$ time-slot, at most $k$ groups require further probing and there are  approximately $\binom{\left \lceil ek \right \rceil}{k}$ combinations possible. Thus, for the frequency-multiplexed GT, the required feedback is $N_t^{opt} \times \log\binom{\left \lceil ek \right \rceil}{k}$ bits. Using upper bound for the binomial coefficient, we have \vspace{-0.2cm}
\begin{equation}\vspace{-0.2cm}
    \binom{\left \lceil ek \right \rceil}{k} \leq \Bigg(\frac{e \times\left \lceil ek \right \rceil}{k}\Bigg)^{k}
    \label{ubbc}
\end{equation}Using 
    $e \times\left \lceil ek \right \rceil  <e^2 \times (k+1),$ we can write, \vspace{-0.1cm}
\begin{equation}
    \log\binom{\left \lceil ek \right \rceil}{k} <2k\log e +k \log \Bigg( \frac{k+1}{k}\Bigg).
    \label{fbub}\vspace{-0.2cm}
\end{equation}Thus, the required feedback is upper-bounded as
\begin{equation} \vspace{-0.2cm}
    N_{FB} <  k \log \Big(\frac{n}{k}\Big)\times \Bigg(2+ \frac{1}{\log e }\log \Big( \frac{k+1}{k}\Big)\Bigg).
\end{equation}\vspace{-0.2cm}
\\
\textbf{Case 2: }$N_t^{UB} < N_t^{opt}$  

 In this case, the number of groups at each time-slot is  $G_i^{*}$ as in (\ref{gsiz}). Thus, the required feedback can be computed as:
\begin{equation}
\begin{aligned}
   N_{FB}= N_t^{UB} \times \log \binom{G_i^{*}}{k} \hspace{2.5cm}&\\
   \leq N_t^{UB} \times \log \bigg(e  \Big(\frac{n}{k}\Big)^{\frac{1}{N_t^{UB}}}\bigg)^{k}\hspace{1.2cm}&\\
    =N_t^{UB}\times k\log e + k \log \Big(\frac{n}{k}\Big). \hspace{1.1cm}
\end{aligned}
\end{equation}Thus, in both cases, our proposed frequency-multiplexed GT asymptotically requires a  feedback of $O\big(k \log (\frac{n}{k})\big)$ bits which is essentially the same as the feedback required for Hwang's GBS. However, an important distinction is that with frequency multiplexed GT, there are $N_t=min(N_t^{opt},N_t^{UB})$ feedback instances each providing $O\big(\frac{k}{N_t} \log (\frac{n}{k})\big)$ bits of information, whereas with Hwang's GBS, there are $O(k \log \big(\frac{n}{k}\big))$ feedback instances of 1-bit. Clearly, frequency multiplexed GT requires $O(k)$ lesser feedback instances.
\section{Numerical simulation}

In this section, we present simulation results validating our theoretical results and establishing the utility of the frequency-multiplexed GT. We consider $n=10^{3}$ devices in our simulations. Also, the number of active devices $(k)$ was assumed to range from 2 to 20. While performing simulation, we randomly tagged $k$ out of the $n$ devices as active. During each stage of frequency-multiplexed GT, the sensors which were not eliminated in the previous stages are randomly assigned to a group based on  (\ref{subsz}) and (\ref{gsiz}).

\begin{figure}   
	\centering
	\includegraphics[width=0.65\linewidth,height=45mm]{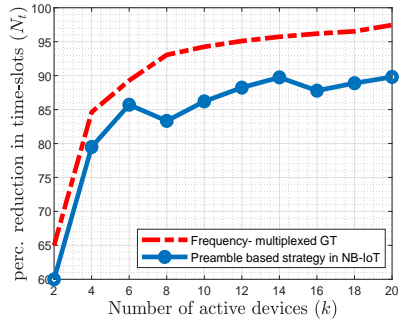}
	\setlength{\belowcaptionskip}{-10pt}
	\caption{Percentage saving in time-slots  for frequency-multiplexed GT and preamble based strategy (NB-IoT)  in comparison to GBS.  Assume $n=10^3$ and $s=48$ (for NB-IoT).}
	\label{fig:tssav}
	\end{figure}

In Fig.\ref{fig:tssav}, we compare the frequency-multiplexed GT strategy described in Section III as well as preamble based access strategy in NB-IoT in terms of the access-delay (number of time-slots needed) w.r.t the Hwang's GBS. For NB-IoT we assumed $s=48$ and a target success probability of 99\% in (\ref{psucpr}). Clearly, we can see a significant reduction in the access delay when we employ the frequency-multiplexed GT strategy instead of GBS. Moreover, as shown in Fig.\ref{fig:tssav}, our proposed frequency-multiplexed GT performs superior to the 3GPP standardized preamble based strategy for NB-IoT.

\begin{figure}   
	\centering
	\includegraphics[width=0.65\linewidth,height=45mm]{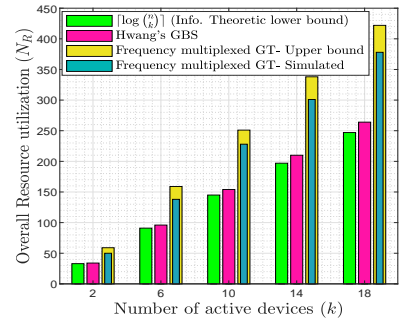}
	\setlength{\belowcaptionskip}{-10pt}
	\caption{Comparison of  total resource utilization when $n=10^{3}.$}
	\label{fig:barp}
	\end{figure}

The variation in overall resource utilization $(N_R)$ as a function of number of active devices $(k)$ for  frequency-multiplexed GT strategy and Hwang's GBS  is shown in Fig. \ref{fig:barp}. The information theoretic lower-bound of $\lceil \log {n \choose k} \rceil$ is also included as a benchmark. Clearly, as expected, Hwang's GBS being the optimal strategy is close to the benchmark.  Our proposed strategy requires more resources which is the price paid for reducing access-delay as shown in Fig.\ref{fig:tssav}.
	
\section{Conclusion}

In this paper, we have proposed a frequency-multiplexed GT strategy for activity detection  using multiple subcarriers in parallel to tackle the access-delay constraints. We have analyzed our proposed scheme in the asymptotic and non-asymptotic regime of $n$ and $k$. We have also derived  the optimal design parameters which minimize the worst-case resource utilization.  It was shown that, in the asymptotic regime of $n$, the frequency-multiplexed GT can deliver order-optimal performance in terms of resource utilization while saving the access-delay by an $O(k)$ factor. Furthermore, we have studied the impact of estimation errors in $k$ on the resource utilization and inferred that even with reasonable estimation errors our algorithm operates efficiently without overwhelming the system resources.  We have also investigated the feedback overhead associated with different schemes of activity detection.

\bibliographystyle{IEEEtran}
\bibliography{THz_scatter}
\end{document}